\documentclass[12pt,a4paper]{article}

\usepackage{epsfig,amsmath,amssymb}

\usepackage{cite}

\def\be{\begin{equation}}
\def\ee{\end{equation}}
\def\lp{\left(}
\def\rp{\right)}
\def\lb{\left[}
\def\rb{\right]}
\def\om{\omega}

\begin{document}

\title{Aspects of spherical shells in a $D$-dimensional background} 
\author{Ernesto F. Eiroa$^{1,2,}$\thanks{e-mail: eiroa@iafe.uba.ar}, 
Claudio Simeone$^{2,3,}$\thanks{e-mail: csimeone@df.uba.ar}\\
{\small $^1$ Instituto de Astronom\'{\i}a y F\'{\i}sica del Espacio, C.C. 67, Suc. 28, 1428, Buenos Aires, Argentina}\\
{\small $^2$ Departamento de F\'{\i}sica, Facultad de Ciencias Exactas y 
Naturales,} \\ 
{\small Universidad de Buenos Aires, Ciudad Universitaria Pab. I, 1428, 
Buenos Aires, Argentina} \\
{\small $^3$ IFIBA, CONICET, Ciudad Universitaria Pab. I, 1428, 
Buenos Aires, Argentina}} 

\maketitle

\begin{abstract}

In this work, shells are mathematically constructed by applying the cut and paste procedure to $D$-dimensional spherically symmetric geometries. The weak energy condition for the matter on the shells is briefly analyzed. The dynamical evolution is studied, and the general formalism for the stability of static solutions is presented. Several examples corresponding to different spacetime dimensions and values of the parameters are considered. 

\end{abstract}

\section{Introduction}

The study of thin shells in general relativity has been developed mainly in the framework of the formalism introduced by Darmois and Israel \cite{daris}. The central tool for analyzing the matter characterization and dynamics of surface layers are the Lanczos equations \cite{daris,mus}, which relate the surface energy-momentum tensor of a shell with  the jump of the extrinsic curvature tensor across it. Apart from cosmological applications, the Darmois--Israel formalism has been applied to highly symmetric configurations, as spherical and cylindrical shells. The linearized stability analysis of spherical shells  was carried out by several authors (see  \cite{sta,eisi11} and the references included in these works).  The formalism was applied to bubbles, shells around stars and black holes, and in the mathematical construction of traversable Lorentzian wormholes (see for example \cite{book,wh,gstars} and references therein). Shells in more than four dimensional backgrounds have been considered in the construction of wormholes \cite{gb06,wh5}. Also, the Darmois--Israel formalism was applied to the collapse of a spherical dust shell into a  Reissner--Nordstr\"om black hole in $D$ spacetime dimensions \cite{gao}. Here we address the characterization and general aspects of the dynamics of spherical shells in backgrounds of $D\geq 4$ dimensions, within the context of $D$-dimensional Einstein gravity and Maxwell electromagnetism. In section 2 we present the general formalism for spherical shells. In section 3 we briefly discuss the weak energy condition in relation with the masses and charges of  the original metrics from which the construction starts. In section 4 we consider a non perturbative treatment of the dynamics and discuss some particular cases (an explicit solution is given in the Appendix). In section 5 we present a general perturbative approach of the dynamics, suitable for a stability analysis of static solutions, and apply it to different spacetime dimensions; explicit results are shown and compared for four, five and six spacetime dimensions.  Finally, the results are summarized in section 6. We adopt usual units such that $c=1$.

\section{Mathematical construction}

We start the construction from two spherically symmetric $D$-dimensional manifolds $\mathfrak{M}_-$ and $\mathfrak{M}_+$ with metrics 
\be
ds_\pm^{2}=-f_\pm(r_\pm) dt_\pm^2+f^{-1}_\pm(r_\pm) dr_\pm^2+r_\pm^2d\Omega_n^2
\ee
($n=D-2$) and boundaries $\Sigma_-$ and $\Sigma_+$. 
We identify the boundaries: $\Sigma_-=\Sigma_+=\Sigma$, and then paste the manifolds $\mathfrak{M}_-$ and $\mathfrak{M}_+$ at the hypersurface $\Sigma$, defined by ${\cal F}(r,\tau)= r-a(\tau)=0$, with $\tau$  the proper time measured by  an observer on the surface. The resulting manifold $\mathfrak{M}=\mathfrak{M}_-\cup\mathfrak{M}_+$ is geodesically complete and the corresponding line element is continuous across $\Sigma$ as long as the coordinates in each side are related by $f_-(a)dt_-^2=f_+(a)dt_+^2$. 
 The induced metric on $\Sigma$ is of course unique and has the form
\be
ds_\Sigma^2=-d\tau^2+a^2(\tau)d\Omega_n^2.
\ee
The joining of the two $D$-dimensional metrics implies a matter shell placed at $r=a$. Associated with this, we have a  jump of the extrinsic curvature $K_i^j$ across the surface, which is related  to the energy-momentum tensor $S_i^j$ on the ($D-1$)-dimensional manifold by the Lanczos equations \cite{mus}
\be
\langle K_i^j\rangle-K\delta_i^j=-8\pi S_i^j,\label{e10}
\ee
 where $ \langle K_i^j\rangle \equiv {K_i^j}_+-{K_i^j}_-$, $ K=\langle \delta _j^i K_i^j\rangle $. The components of the extrinsic curvature at both sides of the joining surface  read 
\be
{K_{\theta_k }^{\theta_k }}_\pm =\frac{1}{a}\sqrt{f_\pm(a)+\dot{a}^{2}},
\ee
\be
{K_{\tau }^{\tau }}_\pm =  -\frac{\ddot{a}+f'_\pm(a)/2}{\sqrt{f_\pm(a)+\dot{a}^{2}}}\ ,\label{e9} 
\ee
where $\theta_k $ ($0\le k\le n$) are the angular coordinates, a prime stands for a derivative with respect to the radius and a dot means $ d/d\tau  $. The isotropy of the configuration leads to a simple diagonal surface energy-momentum tensor with
energy density $ \sigma=-S_\tau^\tau$ and pressure $p=S_{\theta_k}^{\theta_k}$, given by:
\begin{eqnarray}
8\pi \sigma & = & \frac{n}{a}\sqrt{\dot{a}^{2}+f_-(a)}-\frac{n}{a}\sqrt{\dot{a}^{2}+f_+(a)},\label{e11}\\
8\pi p & = & -\lp\frac{n-1}{n}\rp 8\pi \sigma -\frac{\ddot{a}+f'_-(a)/2}{ \sqrt{\dot{a}^{2}+f_-(a)}}+\frac{\ddot{a}+f'_+(a)/2}{ \sqrt{\dot{a}^{2}+f_+(a)}},\label{e12}
\end{eqnarray}
where we replaced $ \sigma  $ in the second equation for simplicity. These equations or any of them plus the conservation equation
\be
\frac{d}{d\tau }(\sigma a^{n})+p\frac{da^{n}}{ d\tau }=0
\label{cons}
\ee
are the starting point for the study of the shell dynamics (there is no problem with starting from a static metric of the embedding, because of the Birkhoff theorem). Note that as long as the condition $f_-(a)>f_+(a)$ is fulfilled, the energy density $\sigma$ is positive. In what follows we will also explore in detail this point.

\section{Weak energy condition}

We will restrict our general analysis to shells of non exotic matter. Normal matter at the shell must fulfill the weak energy condition: $\sigma \geq 0$ and $\sigma+p\geq 0$. If any of these inequalities is violated, the shell would be constituted by exotic matter. In this context, it can be convenient to define 
the auxiliary quantity $\Omega_n=2\pi^{(n+1)/2}/\Gamma[(n+1)/2]$.
Thus, within the Einstein--Maxwell framework, the general form of the function $f(r)$ for a black hole in $D=n+2$ dimensions is (see \cite{tanmy})
\be
f(r)=1-\frac{2{\cal M}}{r^{n-1}}+\frac{{\cal Q}^2}{r^{2(n-1)}}-\frac{\Lambda r^2}{3}
\ee
where ${\cal M}=8\pi G_D m/(n\Omega_n)$ and ${\cal Q}^2=2q^2/[n(n-1)]$ with $m$ and $q$ the ADM mass and charge, and $\Lambda$ is the cosmological constant; $G_D$ is the Newton constant in $D$ dimensions.
Therefore, for static solutions ($a$ = \rm{constant}) the weak energy condition to be satisfied by the energy density and the pressure leads to
\begin{eqnarray}
8\pi \sigma & = & \frac{n}{a}\sqrt{f_-(a)}-\frac{n}{a}\sqrt{f_+(a)}\geq 0,\label{e111}\\
8\pi (\sigma+p) & = & \frac{8}{n}\pi \sigma -\frac{f'_-(a)}{2 \sqrt{f_-(a)}}+\frac{f'_+(a)}{ 2\sqrt{f_+(a)}}\geq 0.
\label{e122}
\end{eqnarray}
Note that while in the case $q_\pm=0$ and $\Lambda=0$ the weak energy condition is trivially satisfied if $m_+>m_-$, this is not the situation in general. Normal matter requires at least $f_-(a)>f_+(a)$. Examples in which the weak energy condition would be violated are easy to find: Consider  a charged bubble ($m_-=0$, $q_-=0$, $m_+>0$, $q_+\neq 0$) in a five dimensional background with vanishing cosmological constant. The condition to be outside the horizon is $8G_5m_+/(3\pi )<a^2+q_+^2/(3a^2)$; but this range includes  $8G_5m_+/(3\pi )<q_+^2/(3a^2)$ (note that this does not imply a naked singularity, because here we are not speaking about a point mass, but about a shell of finite radius). Now, if this is the case we have $f_+(a)>1=f_-(a)$, which means that the energy density would be negative if $a^2<\pi q_+^2/(8G_5m_+)$. Then for given charges and masses there is a lower bound for the possible bubble radius. 

A similar care should be taken in more general cases: if the cosmological constant is (reasonably) assumed to be equal inside and outside the shell, one should always start the analysis from a static solution satisfying the necessary condition 
\be
2{\cal M}_--{\cal Q}_-^2a^{1-n} <2{\cal M}_+-{\cal Q}_+^2a^{1-n} .
\ee 
Besides, one should also ensure $\sigma+p\geq 0$. Now, from the conservation equation (\ref{cons}) it is not difficult to show that 
\be
\sigma +p=-a\sigma '/n;
\ee
in practice, then, for a non trivial form of $f_\pm$, the most simple way to deal with the issue of avoiding exotic matter is to numerically find the intersection of both conditions $\sigma\geq 0$ and $\sigma '\leq 0$. This will be the procedure followed in the stability analysis of section 5.

\section{Nonperturbative approach: examples}

The dynamics of the shell is straightforwardly obtained from the last equations of section 2. Squaring twice and rearranging expression (\ref{e11}), we have
\be
\dot{a}^{2}+V(a)=0,\label{mal}
\ee
 which has the form of the energy conservation of a point particle in a one dimensional problem, with the function $V(a)$ playing the role of a ``potential''.  This potential  has the form 
\be
V(a)=\frac{f_-(a)+f_+(a)}{2}-\lb\frac{n\lp  f_-(a)-f_+(a)\rp}{16\pi a\sigma (a)}\rb ^{2}-\lb \frac{4\pi a\sigma(a)}{n} \rb^{2}.\label{pot}
\ee
In order to obtain an equation of motion for $a$, the dependence of $\sigma $ with the shell radius should be given. The conservation equation (\ref{cons}) leads to $a\sigma'+n(\sigma + p)=0$. Once an equation of state yielding a relation $p=p(\sigma)$ is adopted, we can integrate to obtain
\be
\ln a=-\frac{1}{n}\int\frac{d\sigma}{\sigma+p(\sigma)}+C_1.\label{sol1}
\ee
This integral should be inverted to have $\sigma(a)$. From Eq. (\ref{mal}) one has $da/d\tau=\mp \sqrt{-V(a)}$ (recall that from Eq. (\ref{mal}) the function $V(a)$ must be negative along the allowed range of radii), which once $\sigma(a)$ is known can be integrated giving
\be
\tau=\mp \int\frac{da}{\sqrt{-V(a)}}+C_2.\label{sol2}
\ee
Then this relation must be inverted to obtain the solution $a(\tau)$. In brief: for a given equation of state, Eqs. (\ref{pot}), (\ref{sol1}) and (\ref{sol2}) give the general solution of the problem. Clearly, for a general metric function most equations of state will lead to an analytically non tractable problem. However a qualitative analysis is possible for some physically meaningful examples; in the rest of this section we will restrict the discussion to the case of five spacetime  dimensions.
\begin{enumerate}
\item Case $\Lambda =0$, non charged shell  of dust or non relativistic matter ($p\ll \sigma$) around a charged object ($q_+=q_-=q$). In this situation $\sigma\sim a^{-3}$. We set $M=4\pi a^3\sigma/3$ and the potential reads
\be
V(a)= 1-\frac{8G_5(m_-+m_+)}{3\pi a^2}-\frac{1}{4\pi^2}\lb\frac{8G_5(m_--m_+)}{3M}\rb^2+\frac{1}{a^4}\lp\frac{q^2}{3}-M^2\rp.
\ee
Assuming that both $m_-$ and $m_+$ are positive, then we have a monotonous potential for $q^2/3<M^2$, and a potential with a minimum for $q^2/3>M^2$. In the first case an initially static shell can only collapse, while in the second case an oscillatory motion could, in principle, take place. Note that if $q=0$ no oscillations are possible.
\item Case $\Lambda=0$, non charged shell around a non charged black hole ($q_-=q_+=0$), linear equation of state: $p=\eta\sigma$; we assume $0\leq \eta<1$ in order to allow for the interpretation of $\eta$ as the squared velocity of sound on the shell. In this situation the conservation equation leads to the behavior of the energy density $\sigma (a)\sim ka^{-3(1+\eta)}$, and the potential has the form
\be
V(a)=1-\frac{8G_5(m_-+m_+)}{3\pi a^2}-\frac{1}{4\pi^2}\lb\frac{2G_5(m_--m_+)}{k\pi }\rb^2 a^{6\eta}-\lp\frac{4\pi}{3}\rp^2 \frac{k^2} {a^{6\eta+4}}.
\ee
Because we assume $0\leq \eta<1$, now we have a positive power of the radius; however, all the powers appear with a negative coefficient, so that an unbounded motion is not forbidden.
\item A charged bubble ($m_-=q_-=0$, $m_+=m$, $q_+=q$) of dust or nonrelativistic matter ($p\ll\sigma$).   With the definition for $M$ introduced above, the potential reads
\be
V(a)=1-\frac{8G_5m}{3\pi a^2}+\frac{q^2}{6a^4}-\frac{1}{4\pi^2}\lb\frac{8G_5m-\pi q^2/(2a^2)}{3M}\rb^2-\frac{M^2}{a^4}.
\ee
If the bubble has vanishing charge, the collapse is unavoidable, because the potential is a monotonically increasing function of the shell radius, and $V(a)$ goes to $-\infty$ when $a$ goes to zero. For a non vanishing charge, the presence of the associated positive term in the potential could allow a different behavior.
\item A dust shell ($p=0$) around a black hole, in a cosmological constant background ($\Lambda\neq 0$). The potential for such a model has the form
\be
V(a)= 1-\frac{8G_5(m_-+m_+)}{3\pi a^2}-\frac{\Lambda a^2}{3}-\frac{1}{4\pi^2}\lb\frac{8G_5(m_--m_+)}{3M}\rb^2-\frac{M^2}{a^4}.
\ee
Here the interesting situation is that of a negative cosmological constant, because for $\Lambda <0$ (anti-De Sitter background, no cosmological horizon present) the potential diverges when $a\to\infty$, and an unbounded evolution is then excluded.  
\end{enumerate}
In the Appendix we will consider another example for which we will give an analytical solution for the dynamics under certain approximations.  In the following section, instead, we will present a general procedure for studying the stability of static solutions under perturbations preserving the spherical symmetry. The approach will have the nice feature of being independent of the equation of state for the matter on the shell.

\section{Perturbative treatment: stability}
 
As a less general but physically sound approach to the dynamics of the shell we can consider small perturbations preserving the symmetry around a static solution. Our procedure will then be similar to the treatment in Refs. \cite{sta,eisi11}. As noted in the preceding section, once an equation of state is adopted we can formally take the potential as a function of the shell radius. For a perturbative treatment of the stability of static solutions it is enough with the analysis of the first and second derivatives of $V(a)$ at a radius $a_0$ for which $\dot a=0$. Equilibrium satisfies $V(a_0)=0$ and $V'(a_0)=0$, and stability requires $V''(a_0)>0$. We define the functions 
\be
S(a)=\frac{f_-(a)+f_+(a)}{2},\ \ \ \ \ R(a)=\frac{f_-(a)-f_+(a)}{2}
\ee
and we introduce the relation $M=a^n\Omega_n\sigma $, with $M$ not necessarily a constant but a function of the radius. By defining $\om=n\Omega_n/(4\pi)$ the subsequent expressions can take a simpler form. Then the potential can be written as
\be
V(a)=S(a)-\frac{\om^2R^2}{4}\lp\frac{a^{n-1}}{M}\rp^2-\frac{1}{\om^2}\lp\frac{M}{a^{n-1}}\rp^2.
\ee
The first and second derivatives read
\be
V'(a)=S'-\frac{\om^2R'R}{2}\lp\frac{a^{n-1}}{M}\rp^2-\frac{\om^2R^2a^{n-1}}{2M}\lp\frac{a^{n-1}}{M}\rp'-\frac{2M}{\om^2a^{n-1}}\lp\frac{M}{a^{n-1}}\rp',
\ee
\begin{eqnarray}
V''(a)&=&S''-\frac{2}{\om^2}{\lp\frac{M}{a^{n-1}}\rp'}^2-\frac{2M}{\om^2a^{n-1}}\lp\frac{M}{a^{n-1}}\rp''-\frac{\om^2R^2}{2}\lb{\lp \frac {a^{n-1}}{M}\rp'}^2+\frac{a^{n-1}}{M}{\lp\frac{a^{n-1}}{M}\rp''}\rb\nonumber\\
& &-2\om^2R'R\frac{a^{n-1}}{M}\lp\frac{a^{n-1}}{M}\rp'-\frac{\om^2}{2}\lp\frac{a^{n-1}}{M}\rp^2\lb {R'}^2+RR''\rb.
\end{eqnarray}
In the equations that follow it must be understood that the functions are evaluated at an equilibrium radius $a_0$ and the prime means the derivative with respect to $a_0$. Equilibrium implies $V'(a_0)=0$, from what we can express 
\be
\lp\frac{M}{a_0^{n-1}}\rp'=\frac{\om^2a_0^{n-1}}{2M}\lb S'-\frac{\om^2R'R}{2}\lp\frac{a_0^{n-1}}{M}\rp^2-\frac{\om^2R^2a_0^{n-1}}{2M}\lp\frac{a_0^{n-1}}{M}\rp'\rb\equiv X(a_0).
\ee
Stability implies $V''(a_0)>0$, which with this definition leads to
\be
\frac{2M}{\om^2a_0^{n-1}}\lp\frac{M}{a_0^{n-1}}\rp''+\frac{\om^2R^2a_0^{n-1}}{2M}\lp\frac{a_0^{n-1}}{M}\rp''<Y(a_0)-\frac{2}{\om^2}X^2(a_0),
\ee
where
\be
Y(a_0)\equiv S''-\frac{\om^2R^2}{2}{\lp \frac {a_0^{n-1}}{M}\rp'}^2-2\om^2R'R\frac{a_0^{n-1}}{M}\lp\frac{a_0^{n-1}}{M}\rp'-\frac{\om^2}{2}\lp\frac{a_0^{n-1}}{M}\rp^2\lb {R'}^2+RR''\rb.
 \ee
Second derivatives of $M$ would lead to second derivatives of the energy density. But with the aid of the conservation equation we can write
\be
\sigma''=-\frac{1}{a_0}\lb (n+1)\sigma'+np'\rb.
\ee
This seems to introduce the pressure, and consequently the choice on an equation of state, in the formalism. However, because the expressions above apply to the equilibrium, we can define the parameter $\eta\equiv p'/\sigma'$, with both derivatives evaluated at the equilibrium radius, so that only the lowest order of the expansion of the equation of state appears. Then going back from $\sigma$ to $M$ we obtain
\be
\lp\frac{M}{a_0^{n-1}}\rp''=-\frac{n-1}{a_0}\lb\lp\frac{M}{a_0^{n-1}}\rp'-\frac{M}{a_0^{n}}\rb\lp1+\frac{n}{n-1}\eta\rp,
\ee
\be
\lp\frac{a_0^{n-1}}{M}\rp''=2\lp\frac{a_0^{n-1}}{M}\rp^3{\lp\frac{M}{a_0^{n-1}}\rp'}^2+\frac{n-1}{a_0}\lp\frac{a_0^{n-1}}{M}\rp^2\lb\lp\frac{M}{a_0^{n-1}}\rp'-\frac{M}{a_0^{n}}\rb\lp1+\frac{n}{n-1}\eta\rp.
\ee  
Introducing the definition
\begin{eqnarray}
Z(a_0)&\equiv&\om^2 R^2\lp\frac{a_0^{n-1}}{M}\rp^4{\lp\frac{M}{a_0^{n-1}}\rp'}^2\nonumber\\
& &+\frac{n-1}{a_0}\lb \frac{\om^2R^2}{2}\lp\frac{a_0^{n-1}}{M}\rp^3-\frac{2M}{\om^2a_0^{n-1}}\rb\lb\lp\frac{M}{a_0^{n-1}}\rp'-\frac{M}{a_0^n}\rb\lp 1+\frac{n}{n-1}\eta\rp
\end{eqnarray}
the condition for stable equilibrium can finally be put in the form
\be
Z(a_0)<Y(a_0)-\frac{2}{\om^2}X^2(a_0).
\ee
The subsequent analysis needs in general be carried out numerically, and can be performed in terms of the parameter $\eta $, which can be understood as the square of the velocity of sound on the shell. Then, the preferred range would be $0\le \eta< 1$.  

\begin{figure}[t!]
\centering
\begin{minipage}{.47\textwidth}
\centering
\includegraphics[width=\textwidth]{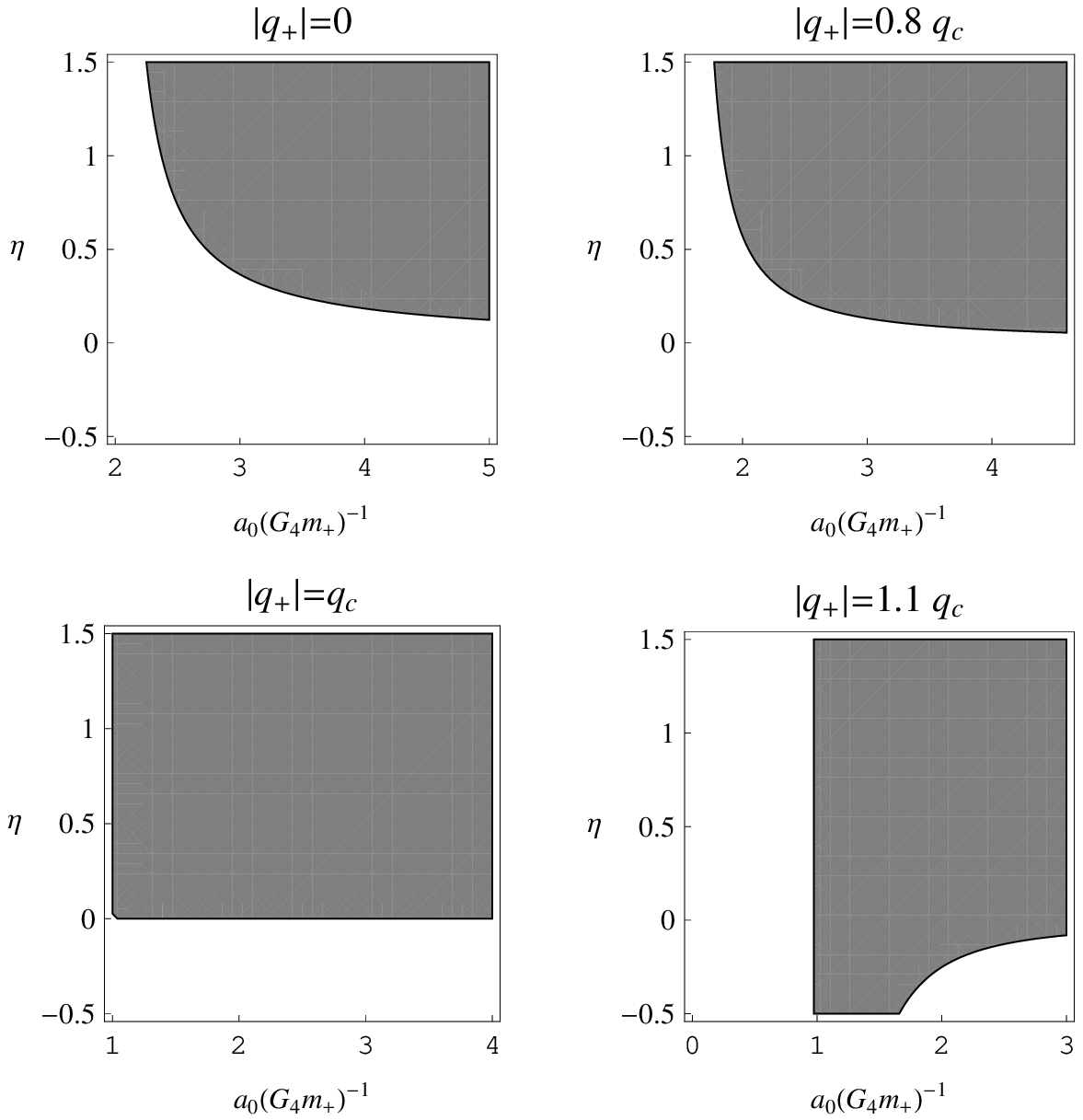}
\caption{Stability regions (grey) for 4-D charged shells around vacuum; $q_c=G_4m_+$. $\mbox{}$}
\label{fig1}
\end{minipage}
\hfill
\begin{minipage}{.47\textwidth}
\centering
\includegraphics[width=\textwidth]{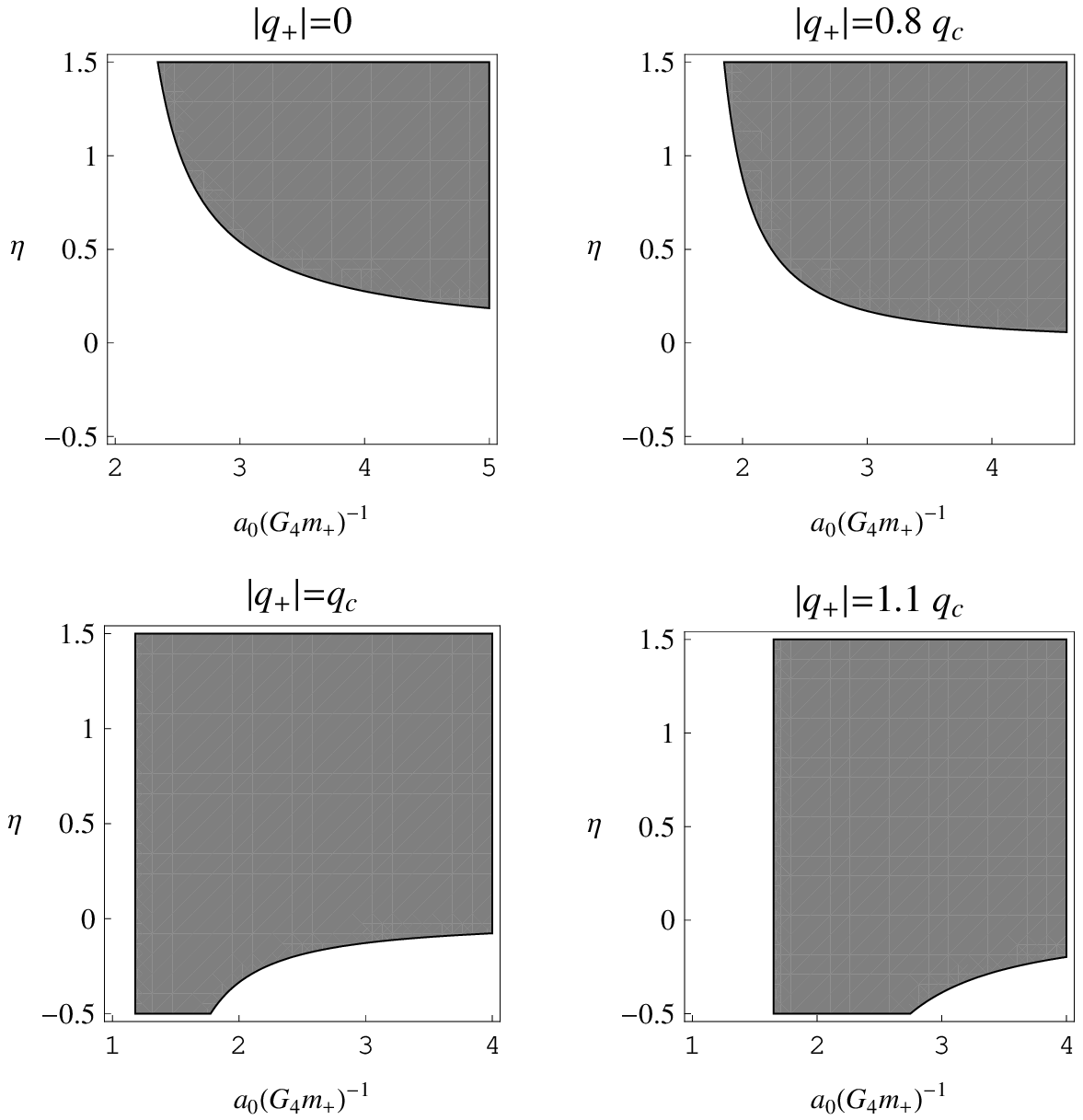}
\caption{Stability regions (grey) for 4-D charged shells around a non charged black hole; $q_c=G_4m_+$.}
\label{fig2}
\end{minipage}
\end{figure}

\begin{figure}[t!]
\centering
\begin{minipage}{.47\textwidth}
\centering
\includegraphics[width=\textwidth]{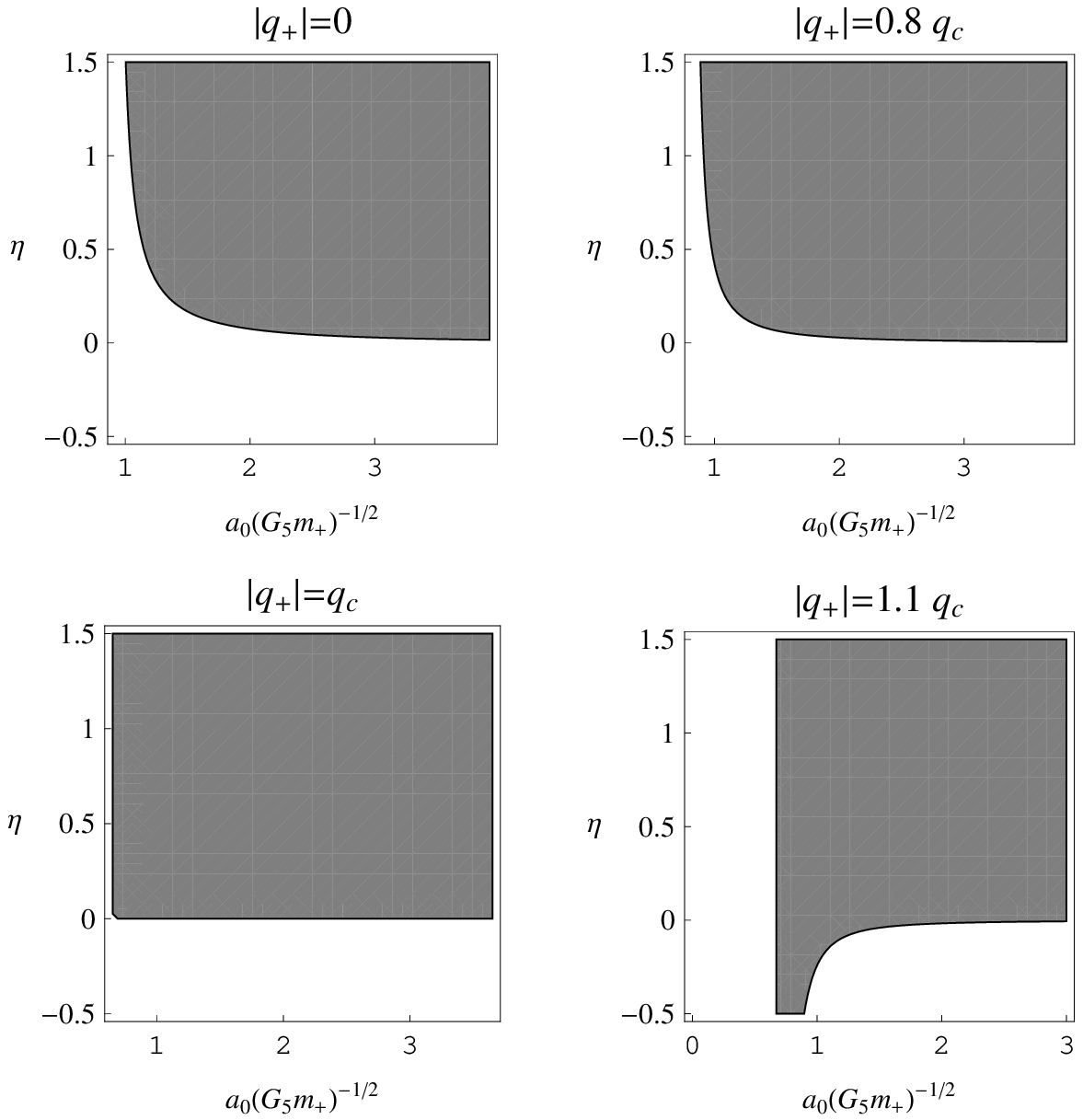}
\caption{Stability regions (grey) for 5-D charged shells around vacuum; $q_c=~4G_5 m_+/(\pi \sqrt{3})$.}
\label{fig3}
\end{minipage}
\hfill
\begin{minipage}{.47\textwidth}
\centering
\includegraphics[width=\textwidth]{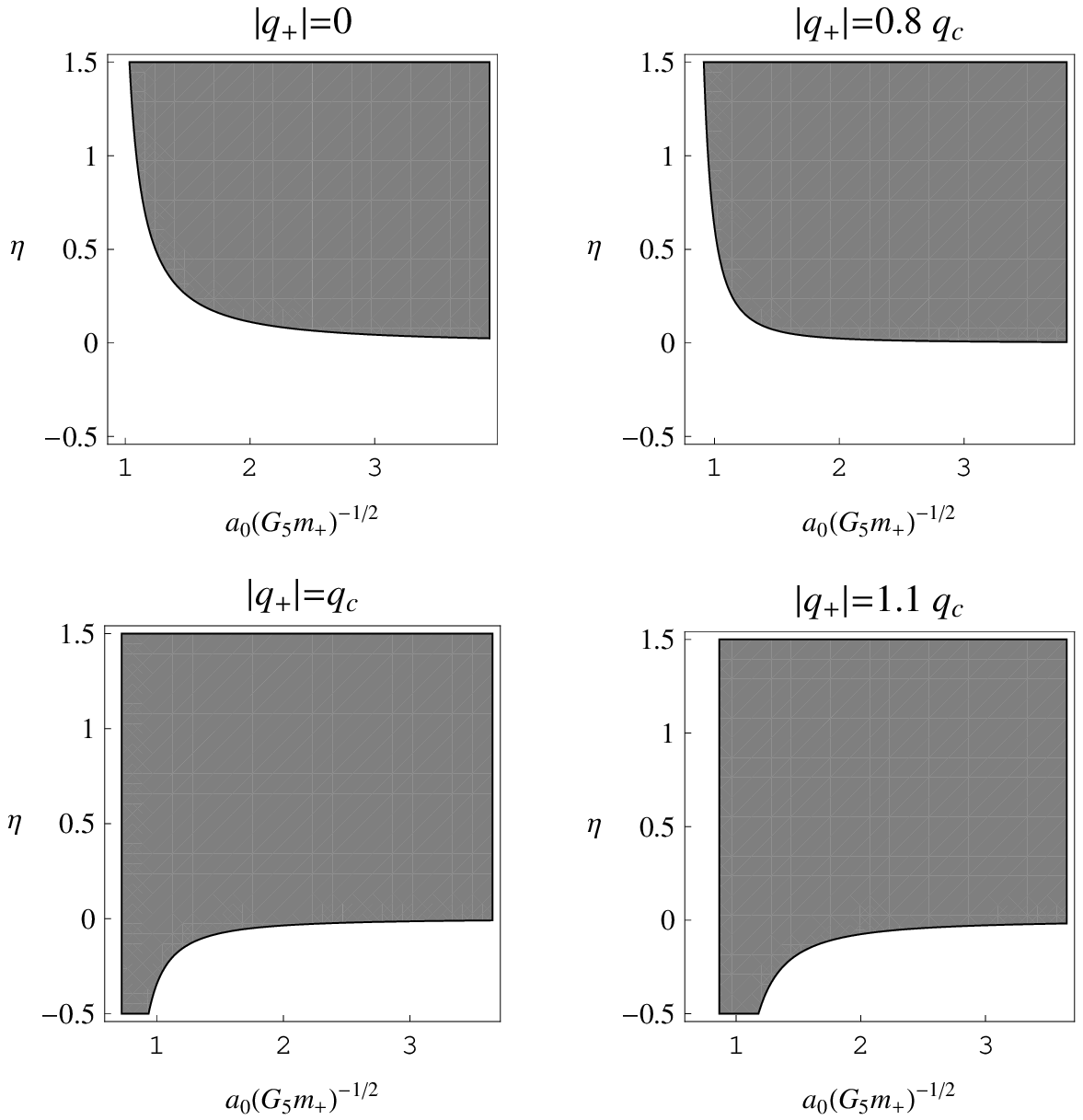}
\caption{Stability regions (grey) for 5-D charged shells around a non charged black hole; $q_c=4G_5 m_+/(\pi \sqrt{3})$.}
\label{fig4}
\end{minipage}
\end{figure}

\begin{figure}[t!]
\centering
\begin{minipage}{.47\textwidth}
\centering
\includegraphics[width=\textwidth]{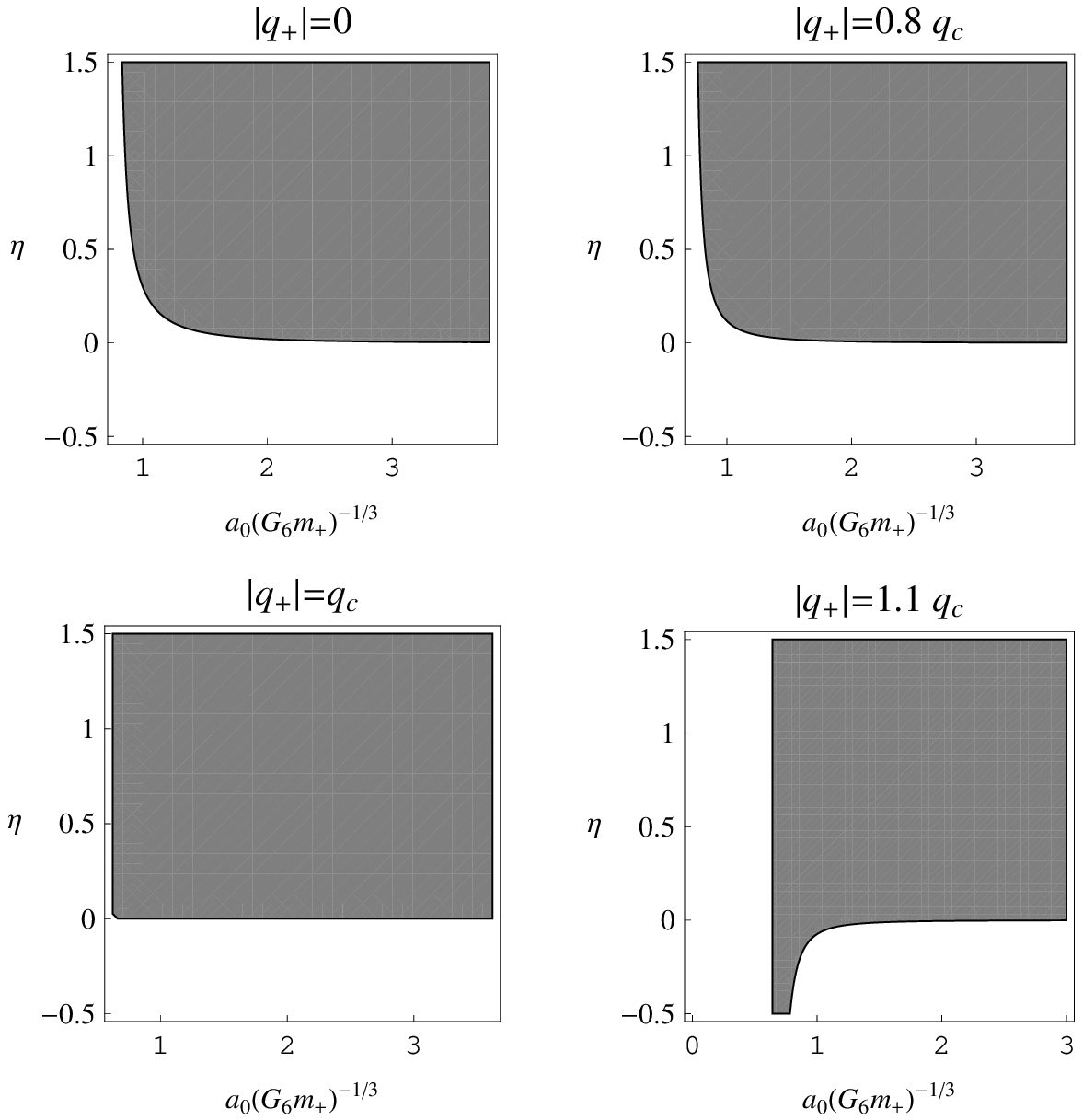}
\caption{Stability regions (grey) for 6-D charged shells around vacuum; $q_c=~3 \sqrt{3}G_6 m_+/(2 \pi \sqrt{2})$.}
\label{fig5}
\end{minipage}
\hfill
\begin{minipage}{.47\textwidth}
\centering
\includegraphics[width=\textwidth]{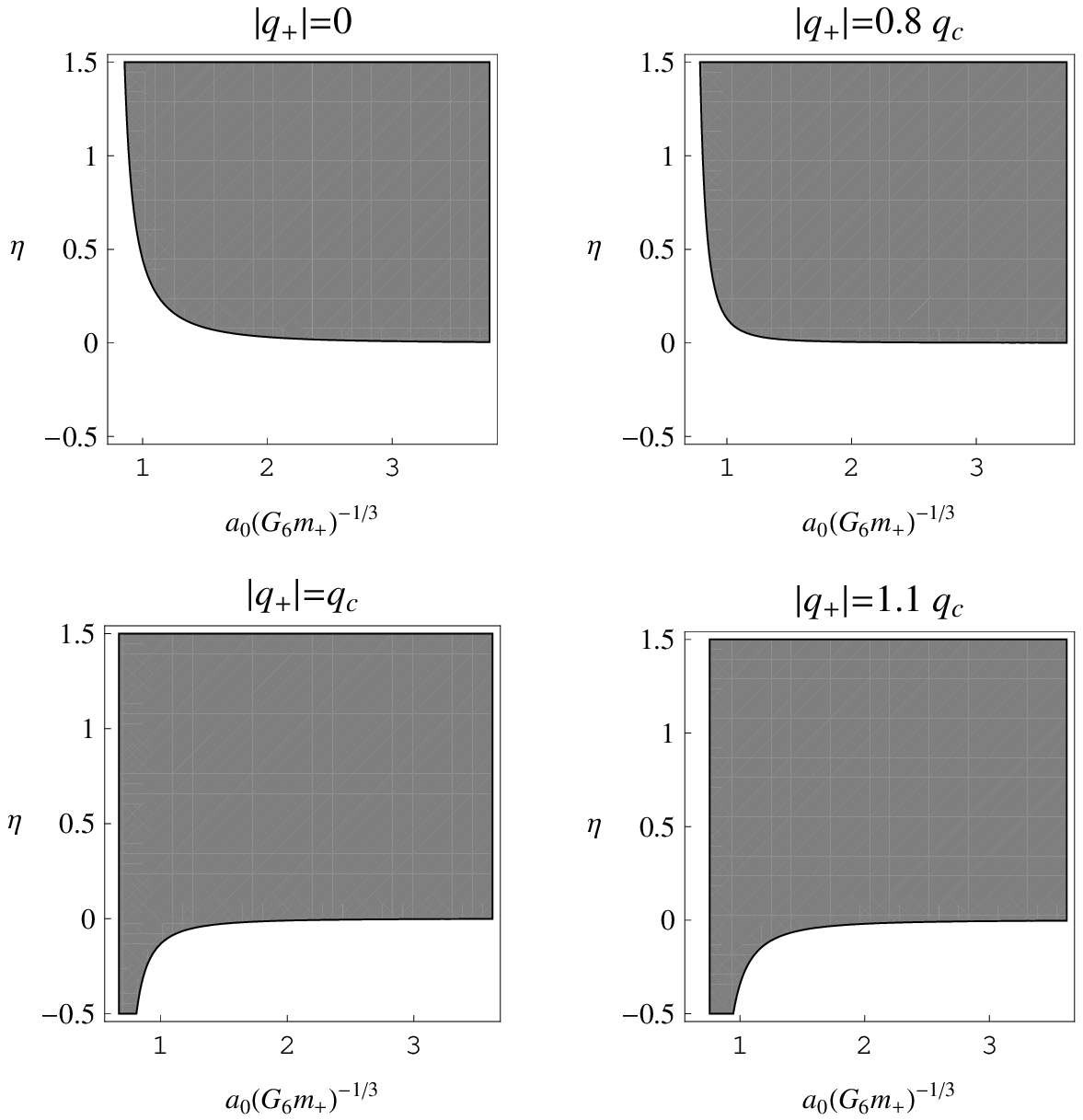}
\caption{Stability regions (grey) for  6-D charged shells around a non charged black hole; $q_c=3 \sqrt{3} G_6 m_+/(2 \pi \sqrt{2})$.}
\label{fig6}
\end{minipage}
\end{figure}

As an application of the formalism, we have studied the stability of charged bubbles ($m_-=0$, $q_-=0$) and charged shells around non charged black holes ($m_-=0.5 m_+$, $q_-=0$) in four, five and six spacetime dimensions. We have set the shell radius beyond the horizon radius of the original outer manifold (so this horizon is removed); which is given by 
\be
r_h= \left(c_1 G_D m_+ +\sqrt{c_1 ^2 G_D^2 m_+^2 -c_2 q_+^2}\right)^{1/(n-1)},
\ee
where $c_1 =8\pi/ (n\Omega _n)$ and $c_2 =2/[n(n-1)]$.  We have restricted the analysis to shells of normal matter, i.e. matter satisfying the weak energy condition. The results are displayed in Figs. 1--6. We have not restricted the graphics to $0\leq\eta<1$, though the results within this range are of more physical interest. We can see that in all cases the stability regions change their shape when the charge reaches the critical value $q_c=G_D m_+ c_1 / \sqrt{c_2}$, from which the horizon in the original outer manifold disappears. For charges below $q_c$, the  regions of stability for bubbles appear to be slightly larger than those for shells around black holes, while the reverse takes place for charges beyond $q_c$; however, for $q\geq q_c$ the stability regions begin at a slightly larger radius as the charge increases. For any number of dimensions, stable bubbles with $\eta<0$ require $q>q_c$, while stable layers with $\eta<0$ around black holes are possible for $q\geq q_c$. The most interesting feature of the results is that the stability regions for fixed $|q_+|/q_c$ are larger as the number of dimensions increases. In particular, as the dimensionality of spacetime increases, for fixed $a_0 (G_D m_+)^{-1/(n-1)}$ stability becomes compatible with smaller values of the parameter $\eta$.   

The procedure above is also valid for wormholes if the outer part of both manifolds is taken, and the signs in the expressions of the energy density (\ref{e11}) and  pressure (\ref{e12}) are suitably changed. In particular, if the two manifolds from which the construction starts are equal copies of a five dimensional spherically symmetric geometry, results including those of Ref. \cite{gb06} can be recovered. In this sense, this section constitutes a generalization of the procedure presented before in that work.

\section{Summary}

We have addressed general aspects of the characterization and dynamics of spherically symmetric shells within the framework of Einstein gravity and Maxwell electrodynamics in $D$-dimensional spacetime. We have applied the Darmois--Israel formalism extended to $D$ dimensions to the mathematical construction of the shells, starting from spherically symmetric geometries associated to, in general, charged black holes with a cosmological constant background. We have discussed the conditions to be imposed on the parameters in order to ensure that the matter on static shells is not exotic, i.e. it satisfies the weak energy condition. Then we have considered the full dynamics preserving the spherical symmetry; we have given the formal general solution of the problem, and discussed some examples by means of the analogy of the equations of motion and the potential of a point particle. Also, an explicit solution is given in the Appendix. Finally, we have presented a formal approach to the study of the stability of static shells under spherically symmetric perturbations; this is given in a concise fashion suitable for the actual application, which in general needs to be carried out in a numerical form. We have applied the formalism to obtain explicit results regarding the stability of charged bubbles and shells around black holes in four, five and six spacetime dimensions. The most interesting result seems to be that as the number of spacetime dimensions increases the stability regions in parameter space become larger.

\section*{Appendix}

Consider the five dimensional metrics with $f_\pm (r)=1-2m_\pm/(\pi r^2)$, in which we have adopted units such as $G_5=3/4$. We assume $p=0$, corresponding to the physically interesting case of a shell of nonrelativistic matter or dust. This assumption implies an energy density with the dependence
\be
\sigma=\frac{k}{a^3}.
\ee
This leads to the equation of motion 
\be
\dot{a}^{2}+1-\frac{m_-+m_+}{\pi a^2}-9\lp \frac{m_--m_+}{8\pi^2 k }\rp ^{2}-\frac{(4\pi k/3)^2}{a^4}=0.\label{motion}
\ee
The time evolution of the shell can be obtained by integrating Eq. (\ref{motion}), which gives the proper time in terms of the shell radius. Defining $\alpha=-1+9(m_--m_+)^2/(8\pi^2 k)^2$ and $\beta=-(m_-+m_+)/\pi$ we have
\be
\tau=\pm\int\frac{a^2da}{\sqrt{\alpha a^4-\beta a^2+(4\pi k/3)^2}}+C.\label{t(a)}
\ee
Inverting the relation given by the solution of this integral yields the shell radius as a function of the proper time. The integral can be expressed in terms of elliptic functions; however, a qualitative analysis can be carried out by recalling that the problem is formally analogous to the dynamics of  a point particle in a potential 
\be
V(a)=-\alpha+\frac{\beta}{ a^2}-\frac{(4\pi k/3)^2}{ a^4},
\ee 
with a null total energy. For $\beta <0$ only an accelerated contraction or a decelerated expansion are possible; however, for $\beta >0$ the possibility of an unbounded accelerated motion exists, for suitable values of the parameters.  It is easy to show that for $\beta >0$ the potential has a maximum for 
\be
a_{\rm m}=\frac{4}{3}\pi k\sqrt{\frac{2}{\beta}},
\ee
so that  $V'(a)<0$ for $a>a_{\rm m}$. Values of the parameters $m_-$ and $m_+$ can be chosen so that $\alpha>0$, $\beta>0$. In this case, depending on  the relation of $\alpha$ and $\beta$ with $k>0$, $V(a_{\rm m})$ can be both positive or negative, while the potential asymptotically tends to the constant $-\alpha<0$ (a simple numerical analysis shows that, taking a scale such that $m_+=1$, for example in the case $m_-=-2$ and $k=0.11$ we have a positive maximum, while for $k=0.1$ the maximum is negative; it is also easy to see that with such values of the parameters the condition $a_{\rm m}>\sqrt{2m_+/\pi}$ is fulfilled). Thus, for such sets of parameters and an initial condition $\dot{a}_i>0$, we have three kinds of evolution: 
\begin{enumerate}

\item If $V(a_{\rm m})$ is negative, for any initial condition  $a_i<a_{\rm m}$ (always beyond the radius $\sqrt{2m_+/\pi}$) the shell radius undergoes a decelerated increase until $a=a_{\rm m}$ is reached, and beyond this point the subsequent evolution is accelerated. 

\item If $V(a_{\rm m})$ is negative but $a_i>a_{\rm m}$, the shell expands with positive acceleration. 

\item If, instead, $V(a_{\rm m})$ is positive, the potential has two positive real roots, and  for $a_i$ above the largest root the shell undergoes an accelerated expansion.

\end{enumerate} 

Because the derivative of the ``potential`` vanishes for $a\to\infty$, in all cases the positive acceleration  decreases as  the radius grows. In brief, the model undergoes an accelerated expansion with decreasing acceleration. This kind of evolution of the shell radius is possible only if $\beta>0$, which implies a negative mass. But this condition  can be fulfilled  without any exotic matter in the $(3+1)$-dimensional manifold resulting from the cut and paste construction: if $m_-<0,\, m_+>0$ and $|m_-|>|m_+|$, then $\sigma >0$ and also $\sigma+p>0$. The exotic matter  --negative mass $m_-$-- is placed under the radius $a$ in the $(4+1)$-dimensional manifold from which we started the construction. Therefore a $(3+1)$-dimensional shell with matter in the form of dust of positive energy density  is compatible with an accelerated expansion. Recall that given the form of the induced $(3+1)$-dimensional metric one could understand the result for the shell as the time evolution of a closed cosmology which, quite interestingly at present, undergoes an accelerated expansion. This accelerated toy ``universe`` would present two positive features:  the unusual kind of matter driving the acceleration is not placed  within $(3+1)$-dimensional spacetime, and the framework in which such kind of evolution is possible is that of classical general relativity. Of course, this is not to be taken too seriously; i.e. we do not pretend that this observation has a phenomenological interest. 

An approximate explicit solution for the time evolution of the model can be obtained for the regime in which the radius $a$ is large enough. More precisely, for $\alpha >0$ and as long as $\alpha a^4-\beta a^2\gg(4\pi k/3)^2$, we can expand 
\be
\lb\alpha a^4-\beta a^2+(4\pi k/3)^2\rb^{-1/2}\simeq\lp\alpha a^4-\beta a^2\rp^{-1/2}\lb 1-\frac{2(2\pi k/3)^2}{\alpha a^4-\beta a^2}\rb.
\ee
Under this approximation the integral form of the time as a function of the radius is
\be
\tau\simeq\int\frac{a\,da}{\sqrt{\alpha a^2-\beta }}-2\lp \frac{2}{3}\pi k\rp^2\int\frac{a^2da}{\lp\alpha a^4-\beta a^2\rp^{3/2}}+C.
\ee
 The integrals are easily calculated, and the result is 
\be
\tau\simeq\frac{1}{\sqrt {\alpha}}\sqrt{a^2-\beta/\alpha}+\frac{2(2\pi k/3)^2}{\alpha^{3/2}}\lb\frac{\alpha}{\beta\sqrt{a^2-\beta/\alpha}}+\lp\frac{\alpha}{\beta}\rp^{3/2}\arccos\lp{\frac{\sqrt{\beta/\alpha}}{a}}\rp\rb+C.
\ee
The approximate evolution of the shell radius is obtained inverting this relation. Note that for very large values of $a$ the evolution is almost linear with time (the zeroth order result is given by the first term, and  is simply $\tau\sim\sqrt{a^2-\beta/\alpha}$), which is consistent with the fact that, as pointed out above, the derivative of the potential tends to zero when $a$ tends to infinity.

\section*{Acknowledgments}

This work has been supported by Universidad de Buenos Aires and CONICET.


\begin{thebibliography}{99}
 
\bibitem{daris} N. Sen, Ann. Phys. (Leipzig) \textbf{378}, 365 (1924); K. Lanczos, Ann. Phys. (Leipzig) \textbf{379}, 518 (1924); G. Darmois, M\'{e}morial des Sciences Math\'{e}matiques, Fascicule XXV  (Gauthier-Villars, Paris, 1927), Chap. 5; W. Israel, Nuovo Cimento B \textbf{44}, 1 (1966); \textbf{48}, 463(E) (1967).

\bibitem{mus} P. Musgrave and K. Lake, Class. Quantum Grav. \textbf{13}, 1885 (1996).

\bibitem{sta} P. R. Brady, J. Louko and E. Poisson, Phys. Rev. D {\bf 44}, 1891 (1991); M. Ishak and K. Lake, Phys. Rev. D {\bf 65}, 044011 (2002); S. M. C. V. Gon\c{c}alves, Phys. Rev. D {\bf 66}, 084021 (2002); F. S. N. Lobo and P. Crawford, Class. Quantum Grav. {\bf 22}, 4869 (2005).
	
\bibitem{eisi11} E. F. Eiroa and C. Simeone, Phys. Rev. D {\bf 83} 104009 (2011).
 
\bibitem{book} M. Visser, \textit{Lorentzian Wormholes} (AIP Press, New York, 1996).

\bibitem{wh}  E. Poisson and M. Visser, Phys. Rev. D \textbf{52}, 7318 (1995); E. F. Eiroa and G. E. Romero, Gen. Relativ. Grav. {\bf 36}, 651 (2004); F.S.N. Lobo and P. Crawford, Class. Quantum Grav. {\bf 21}, 391 (2004); E. F. Eiroa and C. Simeone, Phys. Rev. D {\bf 71} 127501 (2005); M. G. Richarte and C. Simeone, Int. J. Mod. Phys. {\bf D17}, 1179 (2008); E. F. Eiroa, Phys. Rev. D \textbf{78}, 024018 (2008); G. A. S. Dias and J. P. S. Lemos, Phys. Rev. D {\bf 82}, 084023 (2010); J. P. S. Lemos and F. S.N. Lobo, Phys. Rev. D {\bf 78}, 044030 (2008). 


\bibitem{gstars} M. Visser and D.L. Wiltshire, Class. Quantum Grav. \textbf{21}, 1135 (2004);  N. Bili\'c, G.B. Tupper, and R. D. Viollier, JCAP \textbf{02}, 013 (2006); F. S. N. Lobo and A. V. B. Arellano, Class. Quantum Grav. \textbf{24}, 1069 (2007).

\bibitem{gb06} M. Thibeault, C. Simeone and E. F. Eiroa, Gen. Relativ. Grav. {\bf 38}, 1593 (2006).

\bibitem{wh5} M. G. Richarte and C. Simeone, Phys. Rev. D {\bf 76}, 087502 (2007); {\bf 77}, 089903(E) (2008); S. Habib Mazharimousavi, M. Halilsoy and Z. Amirabi, Phys. Rev. D {\bf 81}, 104002 (2010); C. Simeone, Phys. Rev. D {\bf 83}, 087503 (2011).

\bibitem{gao} S. Gao and J. P. S. Lemos, Int. J. Mod. Phys. {\bf A23}, 2943 (2008).

\bibitem{tanmy} F. R. Tangherlini, Nuovo Cimento {\bf 27}, 636 (1963); R. C. Myers and M. J. Perry, Ann. Phys. (NY) {\bf 172}, 304 (1986).


\end{thebibliography}
\end{document}